\documentclass[sigplan,screen]{acmart}
\usepackage[utf8]{inputenc}
\usepackage{graphicx}
\usepackage{listings}
\usepackage{cleveref}
\setcopyright{none}
\usepackage{formaljitterbug}
\usepackage{tikz} 
\usepackage{tikz-cd}
\usepackage{caption}
\usetikzlibrary{positioning}
\usepackage{subcaption}
\usepackage{adjustbox}
\renewcommand\footnotetextcopyrightpermission[1]{}
\acmConference[Dafny Workshop]{}{January 19, 2025}{Denver, Colorado, US}

\begin{document}

\author{Yue Chen Li}
\affiliation{
  \institution{Massachusetts Institute of Technology}
  \country{United States}
}
\email{yuechen.lyc@gmail.com}
\authornote{The author worked on this project during their internship at Amazon Web Services in Boston, USA.}

\author{Stefan Zetzsche}
\affiliation{
  \institution{Amazon}
  \country{United Kingdom}
}
\email{stefanze@amazon.co.uk}
\authornote{Corresponding author}

\author{Siva Somayyajula}
\affiliation{
  \institution{Amazon}
  \country{United States}
}
\email{somayyas@amazon.com}

\title{Dafny as Verification-Aware Intermediate Language for Code Generation}

\renewcommand{\shortauthors}{Li et al.}

\begin{abstract}
Using large language models (LLMs) to generate source code from natural language prompts is a popular and promising idea with a wide range of applications. One of its limitations is that the generated code can be faulty at times, often in a subtle way, despite being presented to the user as correct. In this paper, we explore ways in which formal methods can assist with increasing the quality of code generated by an LLM. Instead of emitting code in a target language directly, we propose that the user guides the LLM to first generate an opaque intermediate representation, in the verification-aware language Dafny, that can be automatically validated for correctness against agreed on specifications. The correct Dafny program is then compiled to the target language and returned to the user. All user-system interactions throughout the procedure occur via natural language; Dafny code is never exposed. We describe our current prototype and report on its performance on the HumanEval Python code generation benchmarks.
\end{abstract}

\maketitle

\section{Introduction}
\label{sec:introduction}

Prompting large language models (LLMs) to generate code from natural language specifications has garnered significant attention from both academia and industry \cite{jiang2024arxiv}. However, such code can either contain bugs or be counterintuitive when compared to the initial specification, despite being confidently presented as correct.

For example, a user might prompt an LLM to generate an efficient Python implementation of the \texttt{Fibfib} sequence (\Cref{fig:fibfibprompt}), which is similar to the classic Fibonacci sequence. It differs in that each term is equal to the sum of the previous three terms instead of the previous two. In \Cref{fig:faultyfibfib} we present Python code that was generated by a state-of-the-art model in response. Despite seeming correct, it contains a subtle error that causes the code to return the wrong value for \texttt{n > 2}. To allow the LLM to recover from its mistake, we provided it with the observation that the test cases in the initial prompt were violated by its synthesised code. In certain instances, this was enough to obtain a correct implementation with its next response. Occasionally, however, it introduced a new bug and, worse, modified the provided test cases to fit its faulty implementation. We will revisit this example later on.

In this paper, we propose to utilise Dafny as an intermediate technology within a code-generating chatbot prototype on the way to higher quality mainstream-language code. Dafny is particularly well-suited for this purpose, because:
\begin{enumerate}
\item It allows for a clear distinction between the specification of a program---derived from a natural language prompt using an LLM---and its implementation, which can again be dialogically derived using an LLM with verification feedback in the loop.
\item The multiple backends supported by the Dafny compiler enable code generation in a variety of mainstream languages, which is ultimately presented to the user, while internally only requiring verification once.
\end{enumerate}

For example, the \texttt{Fibfib} sequence described in (\Cref{fig:fibfibprompt}) can be formalised in Dafny as the recursive function:
\begin{lstlisting}[language=Dafny]
function fibfibFunc(n: int): int
  requires n >= 0
  decreases n
{
  match n
  case 0 => 0
  case 1 => 0
  case 2 => 1
  case _ => fibfibFunc(n-1) + fibfibFunc(n-2) + fibfibFunc(n-3)
}	
\end{lstlisting}
The synthesis problem faced by an LLM then can be specified as the one of finding an imperative implementation and proof hints to fill into the method below:
\begin{lstlisting}[language=Dafny]
method fibfib(n: int) returns (result: int)
  requires n >= 0
  ensures result == fibfibFunc(n)
{...}	
\end{lstlisting}

In the next section, we detail the architecture of the chatbot prototype and explain, in particular, how the use of Dafny itself can be and is opaque to the user. In the following sections, we present some initial benchmarks and the immediate next steps.

\subsection{Related work}

Dafny-guided code generation lies between traditional program synthesis and proof synthesis for math olympiad problems that systems like Google DeepMind’s AlphaProof target for. Work exploring the role of prompt engineering and search-based approaches for the synthesis of Dafny code include that of Misu et.~al.~\cite{misu2024towards} and Branfonbrener et.~al.~\cite{brandfonbrener2024verified}. Here, we utilise few-shot prompting to improve the quality of generated code. Other work leverages Dafny to catch incorrect code, e.g., by automatically synthesizing specifications to check for program correctness \cite{loughridge2024dafnybench}
\cite{mugnier2024laurel}. Lastly, we utilize the idea of consistency between an implementation and its specifications as introduced by Clover \cite{sun2024clover}.

\section{Current Prototype}
\label{sec:currentprototype}

In this section, we provide an overview of a prototype chatbot that is based on Claude Sonnet 3.5 but has been modified to leverage Dafny as an opaque intermediate language for the generation of correct Python code.

\subsection{Higher quality code generation with Dafny}

\begin{figure}
\includegraphics[width=\columnwidth]{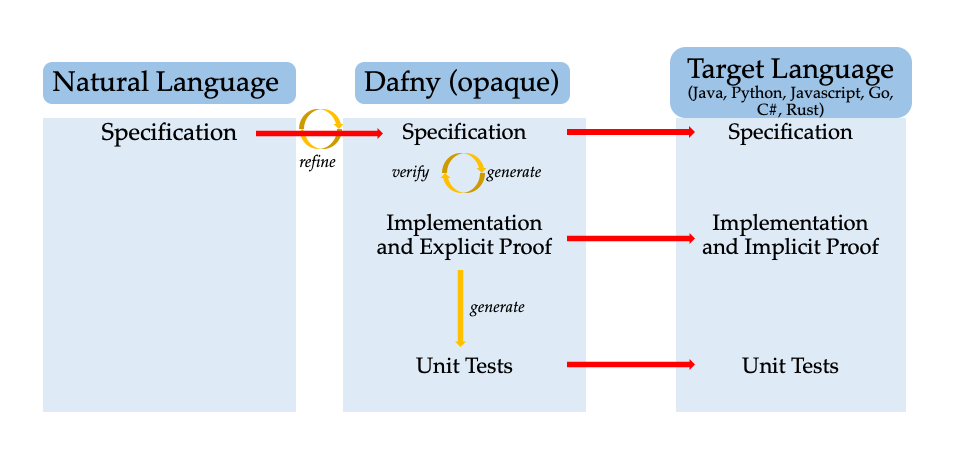}
\caption{Pipeline}
\label{fig:pipeline}
\end{figure}

\Cref{fig:pipeline} depicts the high-level structure of the prototype.

We start on the left, with the natural language input of the user. 
The LLM formalises the prompt in Dafny and returns to the user a natural language translation of the Dafny program that captures the code synthesis task the LLM understands it is supposed to solve. 

The user is able to give their feedback to the formalisation and the natural language conversation loops until both sides agree on a description of the task. This provides a transparent notion of correctness that we can validate a solution against. 

The LLM then generates an attempt at a solution and a proof of its correctness. The Dafny verifier either accepts the candidate solution or rejects it and allows the LLM to refine it or start from scratch again. The process then continues in a loop until a solution appears. This step is completely automated. 

Once the loop has converged, Dafny generates additional unit tests for the solution and then the prototype uses its compiler to translate the solution to the target language. In our case the target is Python, but it could also be Java, etc. The compiler output then may be further optimised and verified for equivalence.

Finally, on the right, the user is presented a specification of the problem in the target language, an implementation, and unit tests. The correctness of the solution is implicit as it was present in Dafny and has been preserved with high assurance by the translation.

\subsection{A demo run}

To illustrate the interactive nature of the pipeline, we prompt the chatbot with a variation of \Cref{fig:fibfibprompt} that contains a typo: in the inductive step, we incorrectly write \texttt{n-4} instead of \texttt{n-3}. 

After a few seconds, the modified chatbot responds with a natural language summary of the underlying Dafny specification of the problem it's been served. Dafny itself is never mentioned in the conversation. There is a slight delay that comes from a loop in the background that ensures only code that is run-time safe and verifiable is returned. At the end of the summary, the LLM points out a few key differences between the initial prompt and the Dafny specification, such as that the input must in fact be non-negative. Most interestingly, it presents a fourth base case \texttt{fibfib(3) == 1} that it says has been implicit in the initial prompt. During the opaque synthesis of the problem specification in Dafny, the verifier seems to have realised that with the typo the recursive definition of \texttt{fibfib} is not well-defined. To solve the issue, the LLM then seems to have synthesised a missing base case, with which the program verified! While the new specification is well-defined, it does not represent our initial intent. Therefore, we respond to the summary by pointing out the typo in our initial prompt.

The LLM comes back with a new specification of the problem which now looks correct, so we accept it and continue with target language code generation. 
After a few moments, we are presented a Python programs that contains an imperative method \texttt{fibfib}, a recursive function \texttt{fibfibFunc}, and a \texttt{Main} method with user provided test cases as well as additional tests that check the equivalence between \texttt{fibfib} and \texttt{fibfibFunc} at runtime (\Cref{fig:prototypesolution}). Under the hood, the LLM has generated an intermediate Dafny program and a proof that it satisfies the agreed on specification, before translating it to Python. The returned Python program now looks good, so we skip on prompting further modifications.
A transcript of the full conversation is given in \Cref{sec:demo-transcript}.

\begin{figure}
\begin{lstlisting}
Please write a Python function `fibfib(n: int)` to efficiently 
compute the n-th element of the FibFib number sequence.

The FibFib number sequence is a sequence similar to the Fibbonacci 
sequence that's defined as follows:
fibfib(0) == 0
fibfib(1) == 0
fibfib(2) == 1
fibfib(n) == fibfib(n-1) + fibfib(n-2) + fibfib(n-3).

The function `fibfib` should satisfy the following test cases:
>>> fibfib(1)
0
>>> fibfib(5)
4
>>> fibfib(8)
24
\end{lstlisting} 
\caption{A user prompting a chatbot to generate an efficient Python implementation of the \texttt{Fibfib} sequence}
\label{fig:fibfibprompt}
\end{figure}

\begin{figure}
\begin{lstlisting}[language=Python]
def fibfib(n: int) -> int:
  if n <= 1: 
    return 0
  elif n == 2:
    return 1 
  else:
    fib_values = [0, 0, 1]
    for i in range(3, n+1):
      next_value = sum(fib_values)
      fib_values = [fib_values[1], fib_values[2], next_value] 
    return fib_values[0]
\end{lstlisting}
\caption{A faulty Python implementation of the \texttt{fibfib} sequence generated by an LLM}
\label{fig:faultyfibfib}
\end{figure}

\subsection{Comparison of native and compiled solutions}

\Cref{fig:faultyfibfib} shows the Python code that was generated directly by an LLM without utilising Dafny in any way. It incorrectly returns \lstinline[basicstyle=\normalsize\ttfamily]{fib_values[0]} instead of \lstinline[basicstyle=\normalsize\ttfamily]{fib_values[2]}.

\Cref{fig:prototypesolution} shows the Python program that was generated by our prototype with Dafny in the loop. The program consists of the following components:
\begin{itemize}
	\item An imperative implementation (\lstinline[basicstyle=\normalsize\ttfamily]{fibfib})
	\item A functional specification (\lstinline[basicstyle=\normalsize\ttfamily]{fibfib_func})
	\item Tests that were provided by the user \\(\lstinline[basicstyle=\normalsize\ttfamily]{Default.fibfib(1) == 0})
	\item Additional tests generated by Dafny \\
	(\lstinline[basicstyle=\normalsize\ttfamily]{Default.fibfib(3) == Default.fibfib_func(3)})
\end{itemize} 
The program is run-time safe and its correctness is implicit through a proof in Dafny.

\begin{figure}[t]
\begin{lstlisting}[language=Python, basicstyle=\tiny\ttfamily]
class Default:
  @staticmethod
  def fibfib(n):
    if n <= 1:
      return 0
    if n == 2:
      return 1
    a, b, c = 0, 0, 1
    for _ in range(3, n+1):
      a, b, c = b, c, a + b + c
    return c

  @staticmethod
  def fibfib_func(n):
    if n == 0 or n == 1:
      return 0
    if n == 2:
      return 1
    return (Default.fibfib_func(n-1) + Default.fibfib_func(n-2) + 
            Default.fibfib_func(n-3))
   
  @staticmethod
  def main():
    assert Default.fibfib(1) == 0, "expectation violation"
    assert Default.fibfib(5) == 4, "expectation violation"
    assert Default.fibfib(8) == 24, "expectation violation"
    assert Default.fibfib(3) == Default.fibfib_func(3), "expectation violation"
    assert Default.fibfib(2) == Default.fibfib_func(2), "expectation violation"
    assert Default.fibfib(1) == Default.fibfib_func(1), "expectation violation"
    assert Default.fibfib(0) == Default.fibfib_func(0), "expectation violation"
 
if __name__ == "__main__":
  Default.main()
\end{lstlisting} 
\caption{The Python implementation of the \texttt{fibfib} sequence generated by our prototype.}	
\label{fig:prototypesolution}
\end{figure}

\subsection{Overview of Dafny code generation strategies}

\paragraph{Prompts provide rules for output generation}

To help the LLM with generating a Dafny specification from natural language, we provide it with a system prompt (\Cref{fig:system-prompt}). The prompt states the LLM's role on a high-level and defines its tasks. It specifies a format for the output and some constraints that enable the easy extraction of the generated code for further processing. To avoid running into common classes of errors, the prompt also instructs the LLM to use certain Dafny syntax, e.g. to use a custom \texttt{Option} datatype to deal with Python functions that could return \texttt{None}. Finally, we instruct it to translate user provided test cases into \texttt{expect} statements that can be checked at runtime.

\paragraph{Few-shot prompts show code organization}

Further guidance to the LLM is given through few-shot prompting (\Cref{fig:fewshotexample}), via multiple input and output pairs that exemplify the high-level instructions of the system prompt. The examples of output Dafny programs share a common structure: they contain a bodyless method that is prefixed with an explanatory docstring and annotated with the \texttt{\{:testEntry\}} attribute, as well as a \texttt{Main} method that provides a fixture for user-provided test cases.

\paragraph{Iterative improvement with Dafny feedback}

The chatbot iteratively converges (automatically, i.e., without user intervention) on a verified Dafny implementation by feeding verifier feedback (errors and warnings) along with a generation task instance. In the process, the LLM not only generates the Dafny implementation code, but also the inline invariants and assertions required to achieve verification.

\paragraph{Post-processing of generated Dafny code}

After a Dafny implementation that correctly implements an agreed-on specification is generated, we must still post-process it to eventually get a target language implementation.
Before we run the Dafny compiler on the generated code, we utilise Dafny's experimental \texttt{generate-tests} command to generate tests in addition to the ones that were provided by the user. The generated tests provide complete coverage of the method that was annotated with \texttt{\{:testEntry\}} (\Cref{fig:fewshotexample}). Dafny offers a choice between different notions of coverage: \texttt{Block}, \texttt{Path}, and \texttt{InlinedBlock}. We used block coverage after inlining (\texttt{InlinedBlock}), a call-graph sensitive version of block coverage. Whereas user provided tests tend to compare the output of a method to a concrete value (\texttt{fibfib(1) == 0}), generated tests will check the validity of an \texttt{ensures} clause at runtime. Often this results in the generated test comparing the output of a method indirectly in the same way as user-defined tests (\texttt{fibfib(1) == fibfib\_func(1)}). To prevent Dafny at times generating excessively long inputs, we set the \texttt{length-limit} option to \texttt{512}.

\subsection{Overview of prototype interaction}

\paragraph{Specification is refined via natural language}

Initially, the user conversationally guides the LLM to an agreed Dafny specification of the code generation task (ref. the initial segment of \Cref{sec:demo-transcript}) by first specifying the input/output behavior of the program and some example input/output pairs in a format similar to the few-shot prompts. Below, we explain how this is achieved without actually revealing any Dafny code to the user.

\paragraph{Consistency checks help validate translations}

One of the main challenges with using Dafny as a purely opaque intermediate language for high-assurance code generation in a target language like Python is that the user and the LLM need to come to an agreement on a Dafny specification through natural language only. That is, we are aiming to prevent the leakage of Dafny specification and implementation code at all times during the chatbot interaction. To allow a discussion, the chatbot refers to natural language representation of internal Dafny code. To ensure that the natural language translation is, as much as possible, an unambiguous description of the Dafny internals, we employ a consistency check as described in \cite{sun2024clover}. In essence, a natural language description is consistent with a Dafny program, if it allows an LLM to reconstruct from it a second Dafny program that is equivalent to the first. We currently use an LLM to decide whether two Dafny programs are equivalent (rather than verifying their equivalence symbolically).

\paragraph{Preserve opacity by filtering system outputs}

To guide the chatbot towards falling back on the Dafny ecosystem if prompted with a Python code generation task, we provide it with a system prompt that outlines a number of steps to follow in such a case. In particular, we utilise Claude's \emph{function calling} capability to point the chatbot at the right moment to externally implemented functionality that either synthesizes a Dafny specification or synthesizes an implementation that is correct with respect to a specification. To prevent intermediate Dafny representations from being surfaced to the user, we mark such calls with a specific tag and selectively hide its output from the user.

\section{Benchmarking}

We’ve evaluated our early prototype on HumanEval, which is a state-of-the-art benchmark set for Python code generation. 

On a set of 164 tasks, the prototype converged to a candidate solution in 144 cases, of which 127 passed all test cases. Overall, this leaves us with a pass rate of 77\%, which is lower than the 86\% for native Python code generation with Claude Sonnet 3.5. However, in a more realistic scenario one falls back on native Python code generation if the prototype does not converge. This would result in a pass rate of $\sim$88\%, a slight improvement over 86\%.
 
The results are promising because there is almost no training data available for Dafny: at the time of writing, there are over 15 million repositories on GitHub that contain Python code, but only 409 repositories with Dafny code. It's worth noting that after moving from Claude 3.0 Sonnet to the newer model Claude 3.5 Sonnet, we saw a dramatic increase in the syntactic correctness of Dafny code generated, which previously accounted for a large class of errors. There are plenty of opportunities to improve these results, for example by synthesizing Dafny code for training and fine-tuning.

\subsection{Most failed tasks have explainable causes}

Most of the 37 cases in which the prototype failed to return a Python program that passes all tests have explainable causes. For around half of the cases, the prototype failed to converge towards a candidate solution, primarily due to a failure to produce verifiable code. If the prototype successfully converged towards a candidate solution, but which then did not pass all tests, the failure was only half of the time due to an algorithmically wrong implementation. Often it instead was due to either too ambiguous a prompt or failing interoperability with native Python code.

\section{Future Work}
\label{sec:future_work}

In this section, we briefly discuss immediate points of improvement for the current prototype.

\subsection{Improving the readability of compiled code}

An issue that we particularly struggled with is that the Python code produced by Dafny's compiler tends to not be entirely idiomatic and requires a Dafny runtime. A possible solution could be to write a more idiomatic Python compiler for Dafny, something we did not attempt at. Another possible solution one can imagine is that the code is post-processed to increase its readability, using heuristics or even an LLM. To keep the correctness guarantees, such a transformation would require an additional check for semantic equivalence. In our chatbot implementation, we post-processed the code with an LLM, but did not verify for semantic equivalence. In \Cref{fig:postprocess} we present an example of such a transformation. For benchmarking, we used the raw output of Dafny's backend, up to some targeted interventions that allowed us to interact with native Python code.

\subsection{Resolving interoperability issues}
\label{sec:interop_issues}

Another inconvenient consequence of depending on the Dafny runtime is that the generated Python code is not always interoperable with native Python code. 

Consider for example the following simple Python program that returns a list with mixed types: 

		\begin{lstlisting}[language=Python]		
def foo():
  return [3.14, "bar", None]			
		\end{lstlisting}

In Dafny, a list is only allowed to contain elements of one type. A program as the one above thus needs to be modelled with the help of a custom datatype:
		\begin{lstlisting}[language=dafny]		
datatype R = A(real) | B(string) | None

method foo() returns (l: seq<R>) {
  l := [A(3.14), B("bar"), None];
}		
		\end{lstlisting}

While this Dafny program successfully compiles to Python, the backend will generate Python classes for it that do not directly compare to the native Python equivalent (\Cref{fig:compiler_seq}). We often ran into this issue during benchmarking, since the test suite was set up in a way that it expects candidate solutions to operate over native Python types. We were able to circumvent some of the issues with the help of custom written wrappers and embeddings, but this solution clearly won't scale. We believe that this particular issue is one of the main hurdles of the overall approach presented here.

\bibliographystyle{ACM-Reference-Format}
\bibliography{bibliography}

\newpage

\appendix

\section{Figures}

\begin{figure}[H]
\begin{lstlisting}
You are an expert Dafny programmer. 

Given a python function signature and docstring, please return a 
Dafny method signature using this format: ```dafny\n<YOUR CODE>\n```.

Please use exactly the same method name as the given function name. 
Ensure the method accepts the same number of arguments as the given 
function.

Use the syntax for a Dafny method. Do not use the syntax for a Dafny 
function. If needed, define and use an `Option<T>` type. (...)

Include the original docstring except for the testcases exactly. 
Create a testing harness method `Main()` that contains the testcases 
translated into dafny `expect` statements.		
\end{lstlisting}
\caption{System prompt}
\label{fig:system-prompt}
\end{figure}

\begin{figure}[H]
\begin{subfigure}[]{\columnwidth}
		\begin{lstlisting}[language=Python]
def foo(l):
    """
    foo takes a list of integers as an input.
    It returns the index of the first 13 if it is in the list, and 
    None otherwise.
    >>> foo([3, 1, 4, 1])
    None
    >>> foo([1, 13, 5, 7, 9, 13])
    1
    >>> foo([])
    None
    """			
		\end{lstlisting}
		\caption{Input}
\end{subfigure}
\begin{subfigure}[]{\columnwidth}
		\begin{lstlisting}[language=dafny]
module M {
  datatype Option<T> = None | Some(val: T)

  /*
  foo takes a list of integers as an input.
  It returns the index of the first 13 if it is in the list, and 
  None otherwise.
  */
  method {:testEntry} foo(integers: seq<int>) 
    returns (index: Option<int>) 
  {
    assume {:axiom} false; // YOUR CODE HERE
  }

  method Main() {
    var t0 := foo([3, 1, 4, 1]);
    expect t0 == None;
    var t1 := foo([1, 13, 5, 7, 9, 13]);
    expect t1 == Some(1);
    var t2 := foo([]);
    expect t2 == None;
}}
		\end{lstlisting}
				\caption{Output}
\end{subfigure}
		\caption{Few-shot prompting}
		\label{fig:fewshotexample}
\end{figure}

\begin{figure}[H]
		\begin{lstlisting}[language=Python]		
@staticmethod
def foo():
  l: _dafny.Seq = _dafny.Seq({})
  l = _dafny.SeqWithoutIsStrInference([
      Ret_A(_dafny.BigRational('314e-2')),
      Ret_B(_dafny.SeqWithoutIsStrInference(
            map(_dafny.CodePoint, "bar")
      )),
      Ret_None()
  ])
  return l			
		\end{lstlisting}
\caption{Custom Python classes generated by the Dafny backend}
\label{fig:compiler_seq}
\end{figure}

\begin{figure}[H]
\begin{subfigure}[]{\columnwidth}
		\begin{lstlisting}[language=Python]
@staticmethod
def fibfib(n__):
  result__: int = int(0)
  if (n__) <= (1):
    result__ = 0
    return result__
  if (n__) == (2):
    result__ = 1
    return result__
  d_0_a_: int
  d_0_a_ = 0
  d_1_b_: int
  d_1_b_ = 0
  d_2_c_: int
  d_2_c_ = 1
  d_3_i_: int
  d_3_i_ = 3
  while (d_3_i_) <= (n__):
    d_4_temp_: int
    d_4_temp_ = ((d_0_a_) + (d_1_b_)) + (d_2_c_)
    d_0_a_ = d_1_b_
    d_1_b_ = d_2_c_
    d_2_c_ = d_4_temp_
    d_3_i_ = (d_3_i_) + (1)
  result__ = d_2_c_
  return result__
  return result__
		\end{lstlisting}
		\caption{Input}
		\label{fig:postprocess-input}
\end{subfigure}
\begin{subfigure}[]{\columnwidth}		
		\begin{lstlisting}[language=Python]		
def fibfib(n):
  if n <= 1:
    return 0
  if n == 2:
    return 1
    
  a, b, c = 0, 0, 1
  for _ in range(3, n + 1):
    temp = a + b + c
    a, b, c = b, c, temp
  return c
\end{lstlisting}
\caption{Output}
		\label{fig:postprocess-output}
\end{subfigure}
		\caption{Post-processing Python code generated by the Dafny compiler with an LLM.}
		\label{fig:postprocess}
\end{figure}

\begin{figure}[H]
  \begin{adjustbox}{width=\columnwidth}
\begin{lstlisting}[basicstyle=\tiny\ttfamily]
<USER>
Please write a Python function `fibfib(n: int)` to efficiently compute the n-th element of the FibFib 
number sequence. 

The FibFib number sequence is a sequence similar to the Fibbonacci sequence that's defined as follows: 
fibfib(0) == 0 
fibfib(1) == 0 
fibfib(2) == 1 
fibfib(n) == fibfib(n-1) + fibfib(n-2)+ fibfib(n-4).

The function fibfib` should satisfy the following test cases:
>>> fibfib(1) 0 
>>> fibfib (5) 4 
>>> fibfib(8) 24

<CHATBOT>
Thank you for providing the task. You've asked me to write a Python function fibfib(n: int) to 
efficiently compute the n-th element of the FibFib number sequence. You've also provided the 
definition of the sequence and some test cases. I'll proceed with designing a specification for 
this function.

The fibfib method takes a non-negative integer n as input and returns the nth term of a modified 
Fibonacci-like sequence. This sequence is defined such that the first two terms are 0, the next 
two terms are 1, and each subsequent term is the sum of the previous term, the term before that, 
and the term four steps back. The method guarantees that for any valid input, it will return the 
correct term of this sequence as calculated by the recursive function fibfibFunc.

The specification is very similar to the original prompt, but it adds a few key details:
1. It explicitly states that the input must be non-negative.
2. It clarifies that the first four terms of the sequence are [0, 0, 1, 1], which was implicit in 
the original definition.
3. It mentions the use of a recursive function (fibfibFunc) to calculate the result, which implies 
an inefficient but correct implementation.

Are you satisfied with this specification, or would you like to make any changes?

<USER>
Oops, I made a mistake. fibfib(n) should be equal to fibfib (n-1) + fibfib (n-2) + fibfib (n-3) 
instead.

<CHATBOT>
Thank you for pointing out the mistake. I apologize for not catching that earlier. Let's update the 
function definition with the correct formula. I'll go back to step 1 and create a new specification 
based on this correction.

Thank you for the correction. I'll now proceed with designing a new specification for the 
fibfib(n: int) function with the updated definition:

The FibFib number sequence is now defined as follows: fibfib(0) == 0 fibfib(1) == 0 fibfib(2) == 1 
fibfib(n) == fibfib(n-1) + fibfib(n-2) + fibfib(n-3) for n > 2

Let me create a new specification based on this corrected definition.

The fibfib method calculates a variation of the Fibonacci sequence for a given non-negative integer 
n. It returns an integer result that matches the value of the fibfibFunc for the same input. The 
fibfibFuncdefines this sequence as for the first two terms, 1 for the third term, and for subsequent 
terms, it's the sum of the three preceding terms in the sequence. This function is guaranteed to 
terminate because the value of n decreases with each recursive call.
  
This specification aligns closely with your corrected prompt. The main differences are:
1. It explicitly states that the input must be non-negative.
2. It mentions the use of a recursive function (fibfibFunc) to calculate the result, which implies 
a correct but potentially inefficient implementation. 
3. It guarantees that the function will terminate, which wasn't mentioned in the original prompt.

Are you satisfied with this updated specification, or would you like to make any further changes?

<USER>
Looks good!

<CHATBOT>
Thank you for confirming that the specification looks good. I'm glad it meets your requirements. 
Now, I'll proceed to generate an implementation based on this specification.		

Here's the Python implementation of the fibfib function, along with the recursive fibfibFunc and 
test cases: 

(...)

This implementation includes:
1. An efficient iterative fibfib function that calculates the nth FibFib number.
2. A recursive fibfibFunc that serves as a reference implementation.
3. A set of test cases in the Main method, including the original tests and additional ones.

Are you satisfied with this Python implementation? If you have any questions or would like any 
modifications, please let me know.

<USER>
Yes!
\end{lstlisting}
    \end{adjustbox}
\caption{Transcript}
\label{sec:demo-transcript}
\end{figure}

\end{document}